# A Logic-Reuse Approach to Nibble-based Multiplier Design for Low Power Vector Computing

Md Rownak Hossain Chowdhury, Mostafizur Rahman
Division of Energy, Matters and Systems, University of Missouri-Kansas City (UMKC)
Kansas City, MO, US
{rhctmc, rahmanmo} @umkc.edu

*Abstract*— **Vector multiplication is fundamental operation for AI acceleration, responsible for over 85% of computational load in convolution tasks. While essential, these operations are primary drivers of area, power, and delay in modern datapath designs. Conventional multiplier architectures often force a compromise between latency and complexity: high-speed array multipliers demand significant power, whereas sequential designs offer efficiency at the cost of throughput. This paper presents a precompute–reuse nibble multiplier architecture that bridges this gap by reformulating multiplication as a structured composition of reusable, nibble-level precomputed values. The proposed design treats each operand as an independent low-precision element, decomposes it into fixed-width nibbles, and generates scaled multiples of a broadcast operand using compact shift-add logic. By replacing wide lookup tables and multiway multiplexers with logic-based precomputation and regular accumulation, the architecture decouples cycle complexity from gate delay. The design completes each 8-bit multiplication in two deterministic cycles with a short critical path, scales efficiently across vector lanes, and significantly reduces area and energy consumption. RTL implementations synthesized in TSMC 28 nm technology demonstrate up to 1.69x area reduction and 1.63x power improvement over shift-add, and nearly 2.6x area and 2.7x power savings compared to LUT-based array multipliers at 128-bit scale.**

## I. INTRODUCTION

Multiplication is a fundamental operation in accelerator datapaths and increasingly dictates overall system efficiency [1]. In contemporary low-precision workloads such as 8-bit inference [2], convolution [3], and SIMD execution [4], throughput is sustained by replicating multiplier units across parallel vector lanes. While reducing arithmetic precision lowers per-instance bit width, it does not proportionally reduce the architectural overhead of replication [5]. As vector width increases to meet throughput demands, the cumulative area, routing complexity, and switching activity of multiplier arrays grow rapidly [6]. Consequently, the structural organization of multiplication becomes a dominant factor in determining accelerator scalability and energy efficiency [7]. This scaling pressure highlights a fundamental trade-off in multiplier design: achieving low latency requires aggressive hardware expansion, while reducing hardware cost sacrifices sustained throughput.

Existing multiplier architectures can be broadly categorized into combinational (single cycle) and sequential (multi-cycle) designs[8]. Combinational multipliers, including array and Wallace-tree structures, fully unroll partial-product generation and reduction to achieve minimal latency [9][10]. However, this approach increases combinational depth, routing complexity, and switching activity as operand width or parallelism grows. Sequential designs such as shift-add and Booth reduce hardware cost by reusing arithmetic resources across cycles, but their iterative execution directly limits throughput in highly parallel accelerator fabrics [11]. As a result, conventional multipliers occupy opposing extremes of the latency–hardware tradeoff.

This work investigates an alternative design point tailored to low-precision vector workloads. The approach is guided by two observations: (1) low-bit multiplication can be expressed as a structured set of recurring shift-and-add patterns, and (2) accelerator workloads frequently broadcast one operand across many independent vector elements. These properties enable multiplication to be reformulated as a composition of reusable digit-level operations rather than as a monolithic arithmetic block. By decomposing each 8-bit operand into two nibbles and reusing a shared broadcast operand, the computation can be expressed as logic-generated scaled values followed by fixed alignment and controlled accumulation.

To realize this idea, we propose a precompute–reuse nibble multiplier that replaces monolithic arithmetic blocks and deep reduction trees with compact logic-based precompute units. Instead of relying on fully unrolled partial-product generation as in Wallace and array multipliers, or iterative accumulation as in shift-add and Booth designs, the architecture generates scaled values on demand through structured shift-and-add logic at the nibble level. This approach eliminates deep adder trees, avoids wide selection networks, and confines computation to narrow, regular datapath blocks. As a result, combinational depth is controlled, routing complexity is reduced, and scaling with operand replication becomes predictable. Within the same design space, we also present a throughput-oriented LUT-based array multiplier that achieves single-cycle execution, enabling a direct comparison between high-throughput and hardware-efficient design points. The key contributions of this paper are summarized as follows:

- **Throughput-Optimized LUT-Based Array Multiplier:** A single-cycle architecture tailored for vector accelerator fabrics where maximum throughput is required.
- **Precompute–Reuse Nibble Multiplier:** A logic-based architecture that replaces wide lookup tables with compact nibble-level precompute blocks, enabling low-power multiplication.

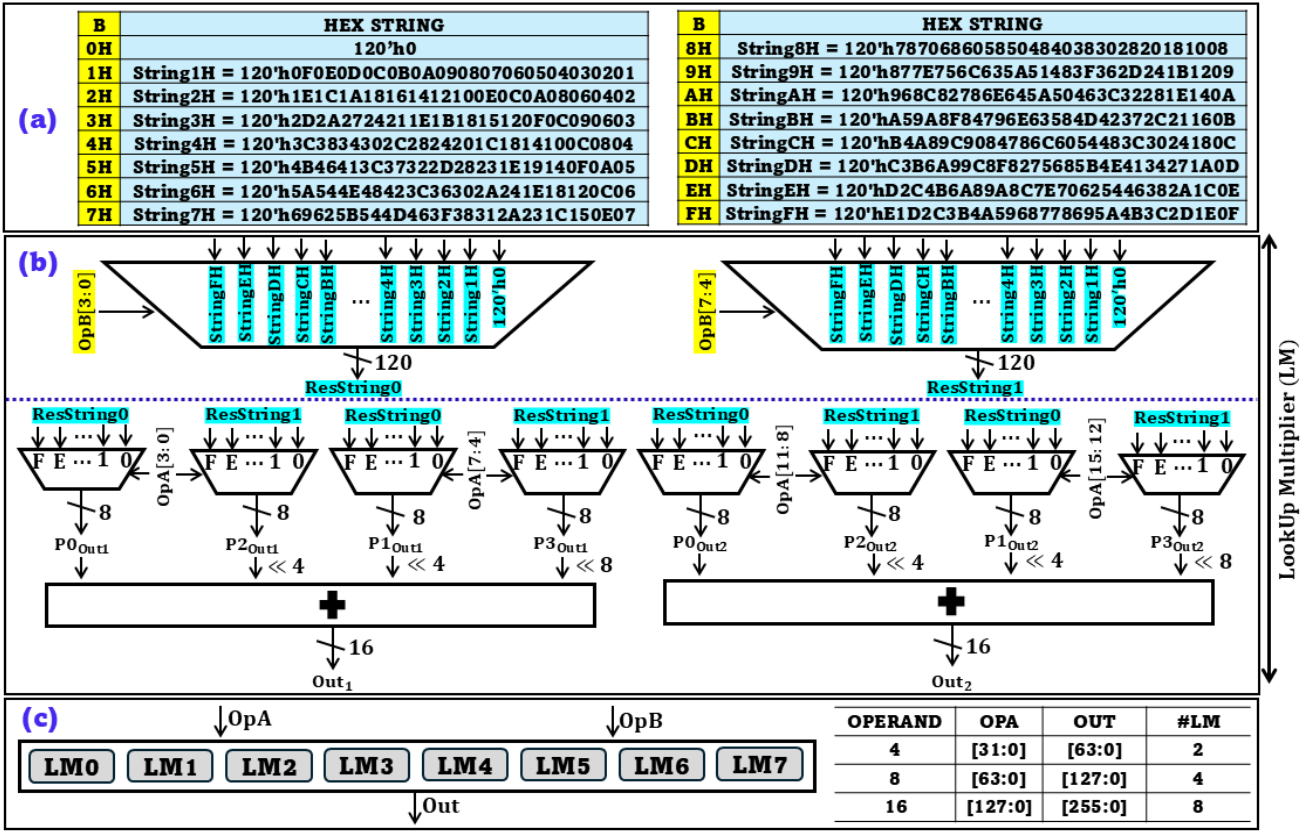

**Fig. 1. LUT-based array multiplier using deterministic hex-string lookup. (a)** Hex-string LUT indexed by operand B nibbles stores precomputed result strings. **(b)** Operation of an LM: nibble-based slice extraction, fixed shifts, and accumulation to produce final product. **(c)** Scalable organization through replication of identical LM blocks for wider array.

- Quantitative evaluation on TSMC 28nm: Comprehensive synthesis in TSMC 28 nm technology across 4, 8, and 16-operand configurations. The nibble architecture achieves up to 1.69x area reduction and 1.63x power improvement over shift-add, and nearly 2.6x area and 2.7x power savings compared to LUT-based array designs.

The remainder of this paper is organized as follows. Section II presents the proposed multiplication architectures, detailing both the throughput-oriented LUT-based array multiplier and the area- and power-efficient precompute–reuse nibble multiplier. Section III provides analytical complexity analysis, cycle comparisons, functional validation, and synthesis-based area and power evaluation across multiple operand configurations. Section IV concludes the paper and discusses the implications of these results for scalable, low-precision accelerator design.

## II. PROPOSED MULTIPLIER ARCHITECTURE

In this section, we first describe the LUT-based array multiplier, which achieves single-cycle execution through hex-string lookup and structured slice composition. We then introduce the precompute–reuse nibble multiplier, which replaces large lookup storage with compact logic-based scaled-value generation to improve hardware efficiency while preserving deterministic execution.

### A. LUT-Based Array Multiplier

The proposed LUT-based Array Multiplier replaces conventional arithmetic multiplication with a structured lookup-and-composition mechanism. Rather than generating partial products and compressing them through multi-level adder trees as in shift-add, Booth, or Wallace architectures, the design removes dynamic partial-product formation from the datapath. As shown in Fig. 1(a), multiplication begins by indexing a compact hex-string lookup table using the nibbles of operand B. For an 8-bit operand, the lower and upper nibbles independently select two precomputed result strings (Algorithm 1, Line 5), each encoding structured multiplication outcomes. These strings are subsequently reused during the processing of operand A, eliminating the need for bitwise product generation at runtime. Thus, this architecture converts multiplication into deterministic selection logic followed by controlled slicing and fixed-position accumulation.

**Algorithm 1. Lookup Multiplier (LM) Operation**

| | |
|---|---|
| | A: 16-bit Operand (4 Nibbles) |
| **Input** | B: 8-bit Operand (2 Nibbles) |
| | Hex String Look-Up Table (LUT) |
| **Output** | Out: 32-bit Product |
| **1:** | $B_0 \leftarrow B[3:0]$; $B_1 \leftarrow B[7:4]$ |
| **2:** | $A_0 \leftarrow A[3:0]$; $A_1 \leftarrow A[7:4]$; $A_2 \leftarrow A[11:8]$; $A_3 \leftarrow A[15:12]$ |
| **3:** | $P0_{Out1} \leftarrow 0$; $P1_{Out1} \leftarrow 0$; $P2_{Out1} \leftarrow 0$; $P3_{Out1} \leftarrow 0$ |
| **4:** | $P0_{Out2} \leftarrow 0$; $P1_{Out2} \leftarrow 0$; $P2_{Out2} \leftarrow 0$; $P3_{Out2} \leftarrow 0$ |
| **5:** | $ResString0 \leftarrow LUT(B_0)$; $ResString1 \leftarrow LUT(B_1)$ |
| **6:** | $P0_{Out1} \leftarrow ResString0 [(8A_0-8):(8A_0-1)]$ // $A_0 \neq 0$ |
| **7:** | $P2_{Out1} \leftarrow ResString1 [(8A_0-8):(8A_0-1)]$ // $A_0 \neq 0$ |
| **8:** | $P1_{Out1} \leftarrow ResString0 [(8A_1-8):(8A_1-1)]$ // $A_1 \neq 0$ |
| **9:** | $P3_{Out1} \leftarrow ResString1 [(8A_1-8):(8A_1-1)]$ // $A_1 \neq 0$ |
| **10:** | $P0_{Out2} \leftarrow ResString0 [(8A_2-8):(8A_2-1)]$ // $A_2 \neq 0$ |
| **11:** | $P2_{Out2} \leftarrow ResString1 [(8A_2-8):(8A_2-1)]$ // $A_2 \neq 0$ |
| **12:** | $P1_{Out2} \leftarrow ResString0 [(8A_3-8):(8A_3-1)]$ // $A_3 \neq 0$ |
| **13:** | $P3_{Out2} \leftarrow ResString1 [(8A_3-8):(8A_3-1)]$ // $A_3 \neq 0$ |
| **14:** | $Out_1 \leftarrow P0_{Out1} + (P2_{Out1} <<4) + (P1_{Out1} <<4) + (P3_{Out1} <<8)$ |
| **15:** | $Out_2 \leftarrow P0_{Out2} + (P2_{Out2} <<4) + (P1_{Out2} <<4) + (P3_{Out2} <<8)$ |

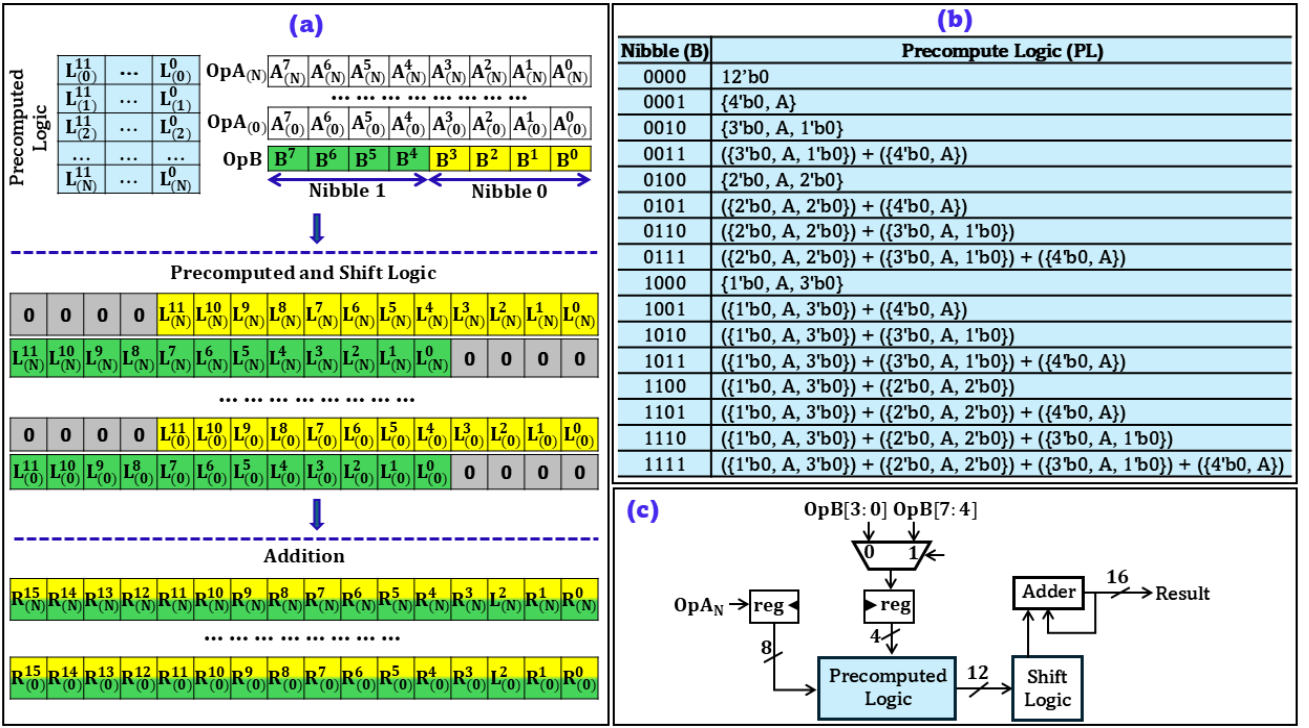

**Fig. 2. Precompute-reuse nibble multiplier using logic-based scaled-value generation. (a)** Overall flow: Input (Vector A, Scalar B) → Precompute and shift processing → Final accumulation. **(b)** Nibble-to-precompute logic mapping, where each 4-bit nibble selects a structured shift-and-add configuration. **(c)** Per-element datapath with operand registration, nibble selection, precompute logic, shift alignment, and accumulation.

Fig. 1(b) presents the internal operation of a single Lookup Multiplier (LM) and corresponds to Algorithm 1 (Lines 6-15). After the two result strings (ResString0, ResString1) are selected from the LUT, each operand A nibble drives a fixed-position multiplexer that selects an 8-bit segment from ResString0 or ResString1 according to the indexing rule defined in Algorithm 1. These extracted segments are organized into two partial product groups to generate output 1 (Lines 6-9) and output 2 (Lines 10-13). The process therefore retrieves pre-encoded result segments through deterministic indexing rather than forming arithmetic partial products. The selected segments are then aligned using fixed shifts and accumulated to produce output 1 (Line 14) and output 2 (Line 15). Thus, the design eliminates partial-product trees and confines computation to structured selection, alignment, and final accumulation.

Fig. 1(c) illustrates how multiple lookup multiplier (LM) instances are arranged to support wider operand configurations. The architecture scales linearly by replicating identical LM blocks, allowing 4, 8, and 16-element modes through simple structural expansion without redesigning the internal datapath. The design achieves single-cycle execution through a fully combinational datapath in which slice selection, alignment, and accumulation occur within the same stage. This constant cycle complexity, however, shifts cost into the selection network where the lookup strings synthesize into large constant logic structures. As vector width grows, these multiplexers and their interconnect increasingly dominate area and power, and the critical path becomes driven by mux depth rather than arithmetic logic. Thus, while the cycle count remains one, delay and energy scale with combinational complexity.

### B. Precompute-Reuse Nibble Multipler

The look-up based array multiplier achieves single-cycle execution by replacing arithmetic trees with structured selection logic. However, its efficiency is limited by large constant tables and wide multiplexing networks that dominate the combinational datapath as operand width scales. To address this limitation, the architecture is refined into a precompute-reuse nibble multiplier that replaces static lookup strings with compact logic-based generation while preserving the core reuse principle. Fig. 2(a) illustrates the core idea of the nibble multiplier. The multiplier receives a vector input A and a scalar input B, where each vector element is processed against the broadcast scalar. The scalar B is internally decomposed into two nibbles, and each nibble selects a predefined precompute-and-shift logic block that operates on the corresponding vector element A. The generated scaled values are then aligned according to nibble position and combined through addition to producing the result. This flow replaces stored lookup strings with structured logic generation while preserving deterministic execution.

**Algorithm 2. Precompute-Resue Nibble Multiplier Operation**

| | |
|---|---|
| **Input** | A [$OpA_0$, $OpA_1$, …, $OpA_{N-1}$]: Each OpA 8-bit |
| | B: 8-bit Operand (2 Nibbles) |
| | Pre-compute Logic (PL) |
| **Output** | R [$R_0$, $R_1$, …, $R_{N-1}$]: Each product 32-bit |
| **1:** | $OpA_{Sel} \leftarrow 0$; $OpB_{Sel(Nibble)} \leftarrow 0$ |
| **2:** | **for** $Idx_A$ = 0 to N-1 **do** |
| **3:** |    Acc ← 0 |
| **4:** |    $OpA_{Sel} \leftarrow OpA\ [Idx_A]$ |
| **5:** |    **for** $Idx_{B(Nibble)}$ = 0 **to** 1 **do** |
| **6:** |       $OpB_{Sel(Nibble)} \leftarrow OpB\ [Idx_{B(Nibble)}]$ |
| **7:** |       Partial ← PL ($OpA_{Sel}$, $OpB_{Sel(Nibble)}$) |
| **8:** |       Acc ← Acc + (Partial << (4 * $Idx_{B(Nibble)}$)) |
| **9:** |    **end for** |
| **10:** |    R [$Idx_A$] ← Acc |
| **11:** | **end for** |

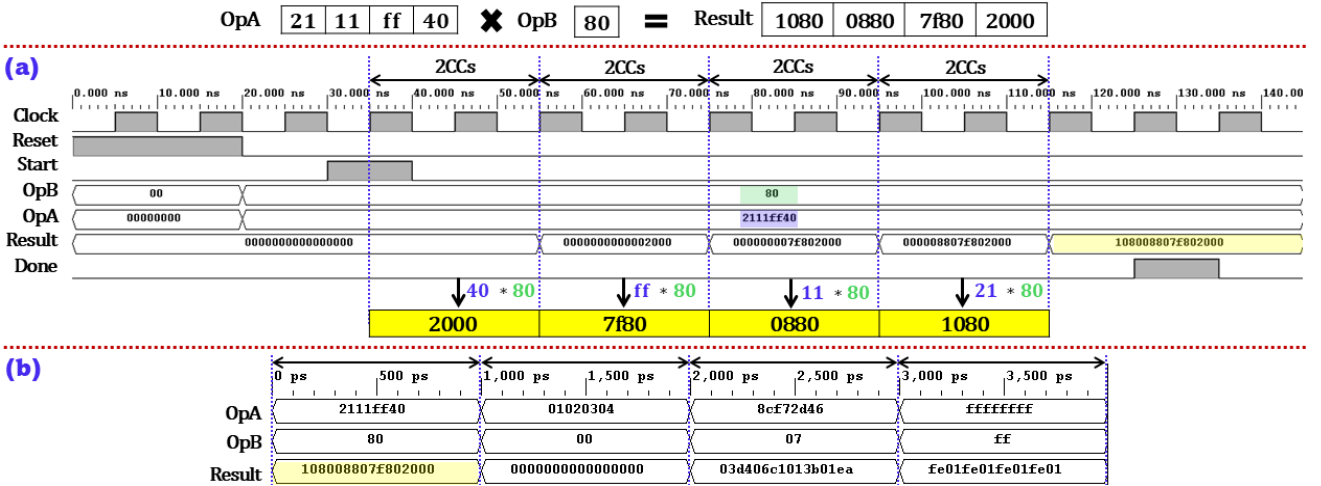

**Fig. 3. Functional verification of 8-operand vector-scalar multiplication. (a)** Precompute-reuse nibble multiplier showing two-cycle execution per vector element with broadcast scalar reuse. **(b)** LUT-based array multiplier demonstrating single-cycle combinational result generation for the same operand set.

Fig. 2(b) details the internal structure of the precompute logic (PL) selected by each nibble of B. Rather than retrieving a stored product, the nibble value activates one of sixteen predefined logic configurations that generate the corresponding scaled version of the vector element A. Each configuration is implemented as a structured combination of fixed shifts and limited additions of A, as summarized in the table. Small nibble values map to single shifted versions of A, while larger values are constructed by summing multiple shifted terms. Fig. 2(c) shows the per-element datapath that implements the nibble multiplier and corresponds to Algorithm 2. For each vector element, the operand A is first selected and registered (Algorithm 2, Lines 2-4), while the scalar B is accessed nibble-by-nibble through the nibble selector (Lines 5-6). The selected A value and current B nibble then drive the precompute logic block, which generates the corresponding scaled partial value (Line 7). This partial value is forwarded to the shift logic, where it is aligned based on nibble position using a fixed shift amount, and accumulated into the running sum through the adder (Line 8). After both nibbles are processed (Lines 5-9), the accumulator holds the complete product for that element, which is written back to the output vector (Line 10).

The architecture supports both sequential and fully unrolled configurations. In the sequential mode, one nibble is processed per cycle, resulting in two cycles for an 8-bit scalar. Each cycle involves only the precompute block, a fixed shifter, and a narrow adder, keeping the critical path short and independent of operand width. In the unrolled mode, both nibbles are evaluated combinationally within a single cycle using duplicated precompute and alignment logic. This refinement significantly improves hardware efficiency compared to the lookup-based array multiplier. The wide lookup strings and large multiplexers are replaced by compact shift-and-add structures that synthesize into small adders, shifters, and registers. The datapath becomes arithmetic-structured rather than selection-dominated, reducing routing congestion and switching activity. As a result, the design achieves a more balanced tradeoff between latency and combinational depth while providing clear area and power advantages.

### III. PERFORMANCE ANALYSIS

This section evaluates the proposed precompute-reuse nibble multiplier against conventional shift-add, Booth (Radix-2), Wallace tree, and LUT-based array architectures. All designs were implemented in synthesizable RTL using Verilog and verified functionally through vector-scalar multiplication testbenches under identical stimulas. Synthesis was performed using a commercial standard-cell flow targeting TSMC 28 nm HPC+ technology, with the detailed experimental parameters summarized in Table 1. All architectures were compiled under identical timing constraints (1 GHz target frequency), voltage (1.05 V), and process corner (FF) to ensure a fair and controlled comparison. The synthesis flow utilized industry-standard CAD tools, and post-synthesis reports were used to extract area (µm²), total power (mW), and cycle latency metrics. Evaluations were conducted across 4, 8, and 16-operand vector configurations. The comparison focuses on three aspects: (i) analytical cycle complexity, (ii) synthesized area scaling, and (iii) total power consumption under different array width.

**Table 1. Experimental Setup**

| Technology | TSMC 28nm |
|---|---|
| Process | HPC+ |
| Metal Stack | 1P8M |
| Voltage (VDD) | 1.05 V |
| Package | Wire Bond |
| Frequency | 1 GHz |

## A. Functional Verification

Fig. 3 validates functional correctness and cycle-level behavior for an 8-operand vector–scalar multiplication. Fig. 3(a) shows the precompute-reuse nibble multiplier executing one vector element every two cycles. The scalar operand B is broadcast and held constant throughout the full operation, while vector elements of A are consumed sequentially. Each element completes after two nibble iterations, producing the expected 16-bit product before advancing to the next vector entry. The waveform highlights the regular timing of the design: fixed two-cycle spacing per element and deterministic output updates, consistent with the nibble-serial accumulation flow.

Fig. 3(b) shows the LUT-based array multiplier under the same stimulus. In this configuration, the multiplication completes in a single combinational step after operands are applied, and the full vector result is produced without iterative cycling. The waveform confirms correct output formation via deterministic lookup and composition, with the result stabilizing within one cycle window. Together, the two waveforms demonstrate that both architectures produce identical functional results while exhibiting distinct execution profiles: the nibble multiplier trades a small, fixed number of cycles for a compact datapath, whereas the LUT-based design achieves single-cycle completion through larger combinational selection logic.

## B. Analytical Complexity and Cycle Comparison

Table 2 summarizes the algorithmic complexity and cycle latency of representative multiplier architectures for 8-bit operands. Sequential multipliers such as shift-add and radix-2 Booth exhibit linear complexity with operand width, reflected in their per-operand latencies of 8 and 4 cycles, respectively. When processing N operands, total latency scales proportionally, resulting in 8N and 4N cycles. These architectures maintain narrow datapaths but incur increasing execution time as operand width grows. On the other hand, combinational designs such as Wallace-tree and array multipliers achieve constant-cycle latency, completing both single and multiple operand cases in one cycle.

However, this constant latency comes with increased combinational depth and wiring complexity. The proposed nibble multiplier, shown in Table 2, occupies tunable middle ground in this space. In sequential mode, latency scales

**Table 2. Analytical complexity and cycle latency comparison of different multiplier architectures**

| Multiplier | Type | Complexity | Latency (CCs) | |
|---|---|---|---|---|
| | | | 1 OpA[#] | N OpA[#] |
| Shift-Add | Sequential | O(W) | 8 | 8N |
| Booth (Radix-2) | Sequential | O(W/2) | 4 | 4N |
| Nibble | Sequential | O(W/4) | 2 | 2N |
| Wallace | Combinational | O(1) | 1 | 1 |
| Array | Combinational | O(1) | 1 | 1 |

W = Operand bit-width; N = Number of Operands; # 8-bit OpA

linearly with operand width as W/4 per operand due to fixed 4-bit decomposition. Consequently, 32 (4 operands), 64 (8 operands), and 128-bit (16 operands) array require 8, 16, and 32 clock cycles, respectively (2N cycles for N parallel 8-bit operands). This scaling makes the cycle cost explicit and predictable while maintaining a short and regular critical path. Unlike traditional combinational multipliers that fix latency to one cycle regardless of width, the nibble architecture trades a small, deterministic increase in cycles for substantially reduced combinational complexity and improved structural regularity.

## C. Performance Evaluation

Fig. 4(a) reports the synthesized area for 4, 8, and 16-operand configurations, normalized to the shift-add baseline. For 4 operands, the shift-add implementation occupies 528.57 µm², while the proposed nibble multiplier requires 463.55 µm² (1.14x of baseline). Booth is comparable at 465.32 µm², Wallace reaches 584.14 µm², and the LUT-based array expands to 806.78 µm². As operand count increases, the differences become more pronounced. For 8 operands, the nibble design occupies 673.60 µm² (1.46x of baseline), compared to 982.42 µm² for shift-add and 1523.72 µm² for the array multiplier. At 16 operands, the nibble architecture scales to 1132.29 µm² (1.69x of baseline), whereas Wallace reaches 2336.54 µm² and the array implementation grows to 2954.20 µm². Across all configurations, the nibble multiplier maintains the smallest area footprint among the evaluated designs. The scaling trend reflects the structural characteristics of each architecture. The nibble multiplier grows through replication of compact shift-and-add datapath blocks, resulting in steady and predictable area expansion. In contrast, designs that rely on wide lookup structures or dense reduction networks experience more rapid growth as operand count increases. Consequently, the proposed logic-based precompute architecture not only achieves lower absolute area but also demonstrates more balanced scaling behavior as vector width expands.

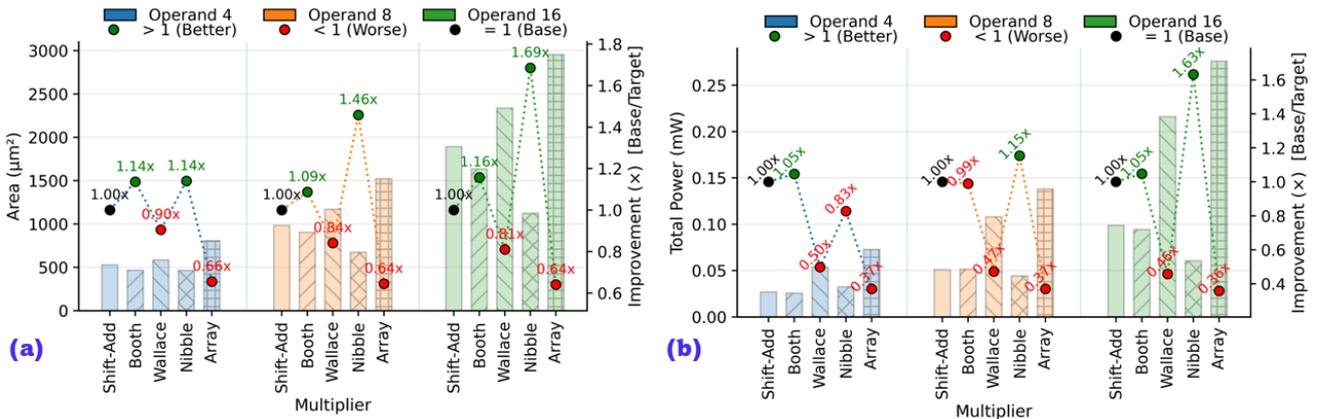

**Fig. 4. Area and power performance across different multiplier architectures for 4, 8, 16-operand configurations.** **(a)** Synthesized area (µm²) with normalized improvement relative to shift-add baseline. **(b)** Total power (mW) with corresponding normalized efficiency across operand scaling.

Fig. 4(b) reports total power in mW for 4, 8, and 16-operand configurations, normalized to the shift-add baseline. For 4 operands, shift-add consumes 0.0269 mW. The nibble multiplier requires 0.0325 mW (0.83× relative improvement), while Booth is slightly lower at 0.0257 mW (1.05×). Wallace and the LUT-based array dissipate 0.054 mW and 0.0727 mW, corresponding to 0.50× and 0.37× efficiency, respectively. At 8 operands, the nibble design consumes 0.0442 mW, improving over shift-add (0.051 mW) by 1.15×, whereas Wallace (0.108 mW) and array (0.138 mW) drop to 0.47× and 0.37× efficiency. The advantage becomes more pronounced at higher parallelism. For 16 operands, the nibble multiplier scales to 0.0605 mW, achieving a 1.63× improvement over shift-add (0.0988 mW). In contrast, Wallace reaches 0.216 mW (0.46×), and the array implementation grows to 0.276 mW (0.36×). The data show that the precompute–reuse architecture not only maintains competitive power at small operand counts but increasingly outperforms other designs as vector width expands. This behavior stems from its compact shift-and-add structure, which limits switching activity and avoids the wide combinational networks that dominate power in alternative implementations.

## IV. CONCLUSION

This paper presented a precompute–reuse nibble multiplier tailored for low-precision workloads, along with a throughput-oriented LUT-based reference design. By decomposing operands into fixed-width nibbles and generating scaled values through compact shift-and-add precompute logic, the proposed architecture replaces deep reduction trees and wide lookup structures with regular, localized datapaths. The design supports both small-constant-cycle sequential execution and single-cycle unrolled operation, explicitly exposing the cycle–delay tradeoff without architectural redesign. Synthesized in TSMC 28 nm technology, the nibble multiplier achieves up to 1.69× area reduction and 1.63× power improvement over shift-add, and nearly 2.6× area and 2.7× power savings compared to LUT-based array multipliers at 16-operand scale. These results demonstrate that nibble-level precompute and reuse provide a scalable, area- and energy-efficient alternative to conventional multiplier architectures for modern AI and domain-specific accelerators.